\documentclass[onecolumn,aps,floatfix,nofootinbib]{revtex4}
\usepackage{latexsym}
\usepackage{mathptm}
\usepackage{amsmath}
\usepackage{amssymb}
\usepackage{graphicx}
\def\beq{\begin{equation}}
\def\eeq{\end{equation}}
\def\beqn{\begin{eqnarray}}
\def\eeqn{\end{eqnarray}}
\newcommand{\f}{\begin{equation}}
\newcommand{\ff}{\end{equation}}
\newcommand{\be}{\begin{equation}}
\newcommand{\ee}{\end{equation}}
\newcommand{\barray}{\begin{array}}
\newcommand{\earray}{\end{array}}
\newcommand{\bea}{\begin{eqnarray}}
\newcommand{\eea}{\end{eqnarray}}

\begin{document}
\title{A real ensemble interpretation of quantum mechanics}
\author{Lee Smolin\thanks{lsmolin@perimeterinstitute.ca}  \\
Perimeter Institute for Theoretical Physics, \\ 31 Caroline Street North, Waterloo, Ontario N2J 2Y5, Canada}

\date{\today}

\begin{abstract}

A new ensemble interpretation of quantum mechanics is proposed according to which the ensemble associated to a quantum state really exists: it is
the ensemble of all the systems in the same quantum state in the universe.  
Individual systems within the ensemble have microscopic states, described by beables.   
The probabilities of quantum theory turn out to be just ordinary relative frequencies probabilities in these ensembles.  
Laws for the evolution of the beables of individual systems are given such that their ensemble relative frequencies evolve in a way that reproduces 
the predictions of quantum mechanics.  

These laws are highly non-local and involve a new kind of interaction between the members of an ensemble that define a quantum state.  These include a stochastic process by which individual systems copy the beables of other systems in the ensembles of which they are a member.  The probabilities for these copy processes do not depend on where the systems are in space, but do depend
on the distribution of beables in the ensemble. 

Macroscopic systems then are distinguished by being large and complex enough that they have no copies in the universe.  They then 
cannot evolve by the copy law, and hence do not evolve stochastically according to quantum dynamics.   This implies novel
departures from quantum mechanics for systems in quantum states that can be expected to have  few copies in the universe.  At the same time, we are
able to argue that the centre of masses of large macroscopic systems do satisfy Newton's laws.  

\end{abstract}

\maketitle

\tableofcontents

\newpage

\section{Introduction}

In this paper we propose a novel interpretation of quantum mechanics that offers new answers to some basic questions about quantum phenomena.  

\begin{enumerate}

\item{} Why do microscopic systems have indefinite values of observables, while macroscopic systems have definite values?

\item{} What is the meaning of the probabilities in quantum physics?

\item{} If the quantum state is associated to an ensemble, where are the members of the ensemble to be found?  

\end{enumerate}

This new interpretation is a theory of beables, and hence solves the measurement problem by asserting that there is a real state of affairs in any quantum system given by the values of the beables.  At the same time, we assert that the quantum state describes an ensemble of individual systems.  

Resolving the measurement problem by means of a theory of beables recalls existing hidden variables theories such as those of 
deBroglie Bohm\cite{dB,Bohm}, Vink\cite{Vink} and Nelson\cite{Nelson}.  However, we aspire to remove an awkward feature of those theories 
which is that the dynamics of the beables of individual systems depend on the wavefunction. In the formulations of de Broglie, Bohm and Vink this is expressed by an equation which asserts that the particle moves in a quantum potential, which is built from derivatives of the wavefunction.  In Nelson's stochastic formulation of quantum theory the osmotic velocity depends on the wavefunction\cite{Nelson,me06}.   This dependence of the dynamics of individual beables on the wavefunction is a characteristic, but most mysterious feature of quantum theory.  

This dependence is awkward because of a principle, which we can call the principle of explanaory closure: {\it anything that  is asserted to influence the behavior of a real system in the world must itself be a real system in the universe.}   It should not be necessary to postulate anything outside the universe to explain the physics within the one universe where we live\footnote{This  argument  and its implications are developed in \cite{robertolee}.}.  This means that the wavefunction must correspond to something real in the world.  In the de Broglie-Bohm interpretation this is satisfied by asserting that the wavefunction is itself a beable.  This results in a dual ontology-both the particle and the wavefunction are real.

But this violates another principle, which is that {\it nowhere in Nature should there be an unreciprocated action.}  This means that there should not be two entities, the first of which acts on the second, while being in no way influenced by it\footnote{Einstein invoked this princniple in a 1921 talk where he objected to  ``the postulation," in Newtonian mechanics, "of a thing (the spacetime continuum) which acts without being acted upon." \cite{Einstein1916}. }.  But this is exactly what the double ontology of  deBroglie-Bohm implies, because the wavefunction acts on the particles, but the positions of the particles play no role in the Schroedinger equation which determines the evolution of the wavefunction.  

A class of interpretations  called "statistical interpretations" aim to overcome the double ontology by asserting that the wavefunction corresponds to an ensemble of systems.  But this falls short of satisfying the principle of explanatory closure unless that ensemble really exists in the world.  It is not sufficient to posit that the wavefunction corresponds to an epistemic ensemble that is defined in terms of our ignorance of the world.   Neither is it acceptable to imagine that there is a spooky way in which "potentialities affect realities."   If the behavior of individual systems is to depend on a wavefuction which corresponds to an ensemble, then the principle of explanatory closure demands that each and every member of that ensemble be a physical system in the universe. 

But if the elements of the ensemble the quantum state represents exist, then perhaps the apparent influence of the wavefunction on the individual entities could 
be replaced and explained by interactions between the elements of the ensemble. By so explaining  the influence of the quantum state on the individual system in terms of a new kind of interaction posited to act between members of the ensemble that the quantum state represents, we satisfy both the principle of explanatory closure and the principle of no unreciprocated action. 

In interpretations in which the ensemble is epistemic it would not make sense to posit interactions amongst members of the ensemble because it would mean that physical particles-the distinguished member of the ensemble that are real- are interacting with shadows that reside only in our ignorance of their true motions.  It would be to have reality depend explicitly on possibility.  But if all the elements of the ensemble are real then there is no barrier to positing new kinds of interactions amongst them.  These interactions are certainly non-local.  But we already have strong reason to assert that any theory of beables that reproduces quantum mechanics must be highly non-local.  

This leaves us with one more question to answer: where do the members of the ensemble corresponding to the ground state of the hydrogen atom reside?  There is a simple, but novel answer that can be given to this question: in the universe.  That is,  {\it  the ensemble corresponding to a hydrogen atom in its ground state is the real ensemble of all the hydrogen atoms in the ground state in the universe.}

The test of this general idea is whether a simple form can be proposed for the interactions amongst the members of the ensemble, that reproduces quantum kinematics and dynamics.  In fact, we will see that a simple form of the interactions, in which the members of the ensemble interact in pairs, suffices.  This simple interaction is that the beables of systems in the ensemble copy each other's states, with a probability that depends on the beables of the systems in the ensemble.    
 
Let us now proceed to make these ideas more concrete.  
This interpretation is based on several simple hypotheses: 

\subsection{Basic hypotheses}

\begin{itemize}

\item{}Quantum mechanics applies to small subsystems of the universe which come in many copies.  Thus, it applies to hydrogen atoms and ammonia molecules, but not to cats or people or the universe as a whole.  Quantum mechanics is hence an approximation to an unknown cosmological theory.

\item{}For each microscopic system, there is an ensemble of systems in the universe with the same constituents, preparation and environment.   A pure quantum
state is a statistical description of one of these ensembles.  The elements of the ensemble will be labeled ${\cal S}_I$ where $I=1,...,N$.

\item{}Each individual microscopic system, ${\cal S}_I$ in the ensemble has two beables.  The first is the value of some observables, which will be denoted $\hat{B}$.  The possible values of 
$\hat{B}$ are indexed by $a=1,...P$ and are denoted $b_a$.   The second beable is a phase $e^{\imath \phi_I}.$  We then assert that the microscopic state of an individual system is the value of the pair of beables, $(a_I, e^{\imath \phi_I}).$ 

\item{}The beables evolve by a discrete and probabilistic rule.  There is a probability in each unit time that each system ${\cal S}_I$ copies the beables of system 
 ${\cal S}_J$.  When this happens, 
 \f
a_I \rightarrow a_J , \ \ \ \ \ e^{\imath \phi_I} \rightarrow e^{\imath \phi_J}
\ff
The probability that this happen will be assumed to be a function of the beables of the two systems as well as the number of systems with the same values
of $\hat{B}$ in the ensemble.  It does not depend on where the members of the ensemble are in the universe.  

\item{}The phases also evolve continuously according to a law that also depends on the distribution of beables in the ensemble.

\item{}We hypothesize that there is a process of {\it phase alignment}, by which the phases of two systems with the same values of $\hat{B}$ evolve to become equal. The dynamics as first posited below conserves the alignment of phases.  After that I present a model for the dynamic alignment of phases.

\item{} Finally, we hypothesize that these ensembles are well mixed by the dynamics just described, so that the probability to make a measurement of the beable $\hat{B}$ on any member of the ensemble and get a particular value, $b_a$,  is given by the relative frequency with which that value appears in the ensemble. 

\end{itemize}

We will expand on the motivation for these hypotheses shortly, and then show how they may be expressed in a form that is equivalent to quantum mechanics.  But what we have said is sufficient to answer the questions with which we opened.

\begin{enumerate}

\item{} {\it Microscopic systems have indefinite values of beables, while macroscopic systems have definite values, because microscopic systems come in many copies, and so are subject to the copy rule, in which they evolve stochastically by copying the beables of members of the ensemble they share.  Macroscopic systems are those that have no copies, anywhere in the universe, hence they are not subject to the copy dynamics.  }

\item{} {\it The probabilities in quantum physics  refer to ordinary relative frequencies in an ensemble of real, existing systems.}

\item{} {\it The members of the ensemble are to be found spread throughout the universe.}

\end{enumerate}

\subsection{More about beables and interactions amongst members of an ensemble}

Before we go on to develop the hypotheses just stated it would be good to revisit some aspects of the motivation.  We begin with the similarities and differences to other theories of beables. 

This proposal shares with hidden variables theories such as deBroglie-Bohm,Vink and Nelson the idea that there are real beables. It shares with Nelson also the idea that pure quantum states correspond to ensembles of individual systems.  However, it differs from all of these interpretations in asserting the  ensemble be physically real, as well as in several other respects.

First, it eliminates the need to pick the configuration space as a beable.  In what follows there is assumed to be a beable observable, $\hat{B}$ but its choice is inessential.  
That this is possible was shown by Vink\cite{Vink}, by giving a deBroglie-Bohm like formulation for a general choice of beables.  Indeed, some of the formal development that follows was inspired by Vink's paper\cite{Vink}. Whether there is a preferred choice for it is a subject for future work. 

Second, we eliminate the double ontology which requires that both the positions of the particles and the wavefunction be beables. This can be criticized as an extravogent hypothesis, which makes the world as ontologically bizzare as interpretations such as many worlds that posit the reality of the quantum state.  

However, the lesson of Nelson's formulation \cite{Nelson}, is that, as explained in \cite{me06}, one cannot succeed in making the whole wavefunction just a derived property of an ensemble, derived from the values of configurations of individual systems.  Given the form of the wavefunction,
\f
\Psi (x, t) = \sqrt{\rho(x.t)} e^{\frac{\imath}{\hbar} S (x,t)}
\ff
it is certainly appropriate to regard the probability density $\rho (x,t)$ as a property of the ensemble and we will do so.  But it is much more challenging to regard the phase $S(x,t)$ as derived from an ensemble.  For one thing, the deterministic evolution equation for the position beable of deBrogle-Bohm theory has the velocity depend on $S(x,t)$.  
%The same is true of the stochastic evolution equation for the forward and backward velocities in Nelson's formulation.  
But, if the rates of change of beables depend on $S(x,t)$ it seems that by our principle of explanatory closure, $S(x,t)$ must also be a beable, or must be determined by beables.  But then this contradicts our second principle of no unreciprocated influence and we find ourselves in trouble.  

To get out of trouble we take a new approach to this conundrum.  We posit that each individual microscopic system has a second beable, which is a phase,
$e^{\imath \phi_I}$.   We also posit that the dynamics forces these to a class of configurations in which they come to depend on the other beables $\hat{B}$.  Hence
$e^{\imath \phi_I} \rightarrow e^{\imath \phi_{a_I}}$, where $a_I$ is the value of the beable $\hat{B}$ in the system $I$.  
Once that is the case the information to determine the  function $S(x,t)$ is to be found distributed in the phase beables of all the individual systems in the ensemble.

An interaction between the beables of individual systems that make up an ensemble that is described by the quantum state may seem a strange and novel idea.  But once we regard the members of the ensemble as all physically real, this is just another interaction between systems in the universe.  Certainly these interactions are highly non-local, but we already know from the experimental tests of the Bell inequalities that any theory of beables that reproduces quantum theory must be highly non-local. 
After all, at one time the idea of an interaction between the Sun and the planets seemed bizzare, because it was a non-local action at a distance.  

Once one accepts this general idea, the next step is to ask how the dynamics of an individual system can depend on the beables of other members of the ensemble in such a way that the predictions of quantum mechanics can be obtained.  This is accomplished in the next section.  We will see that to match the quantum evolution in this picture there must be both a stochastic and a continuous evolution rule.   There is a stochastic process by which {\it one member of the ensemble can copy the beables of another member of the ensemble.}  This stochastic process realizes an idea that the beables of a system we prepare here becomes unpredictably shuffled up with the beables of all the similarly prepared systems in the universe.  There is also a continuous evolution of the phase beables.  Both the stochastic and continuous evolution rules depend on relative frequencies in the ensemble.  

One motivation for the copy rule is the idea that space is an emergent property, as suggested by several proposals for quantum gravity.  If space is emergent, then so is locality. From this perspective, two systems with the same constituents, preparations and environment, but only distinguished by their location in space, may be more closely related than is usually thought.  Indeed, we already know that quantum statistics allows us to give a list of positions where hydrogen atoms in their ground states are to be found, but does not permit us to assert which hydrogen atom is in which position.  If this extends to the level of the beables, then distinct  beable configurations may not be stably located with respect to distinct positions in space.  The whole ensemble of beable states of identical subsystems may then evolve in a way that is not captured by the usual local interactions.  The copy rule is a simple suggestion for this new kind of interaction, which has a simple realization that reproduces quantum mechanics. 
 Other rules might be contemplated, but as we will see the copy rule suffices for our purposes.  
 
 What is nice about the copy rule is that it by itself gives all the dynamics the beables need.  Imagine making a series of measurements of the beable $\hat{B}$ of an atom you hold in your laboratory.  The first measurement is $a_0$.  The second is different, it is $a_1$.  The explanation is not that there was a process by which 
 $a_0$ evolved to $a_1$ but that $a_1$ was copied from another version of that atom somewhere in the universe.  Evolution occurs because on subsequent observations you will be seeing beables copied from the ensemble.  This appears to be like motion as a consequence of the rule that gives the probability for the copy process.  
 Indeed, we will see in Section V that in an appropriate limit in which $\hbar$ can be ignored this can account for classical motion of large bodies.  

In the next section we  put the hypotheses we stated above into precise mathematical form and impose several reasonable physical assumptions on the evolution rules.  In section III we show that a very simple form of the rules then leads to the derivation of Schrodinger quantum mechanics. Section IV presents a model for phase alignment. This is a dynamics for the phases $e^{\imath \phi_I}$ which has a set of degenerate zero energy solutions that impose both phase alignment and Schrodinger dynamics.  There are however issues of the stability of these solutions that remain a subject for further work.   In section V we raise and resolve a question unique to this conception of quantum mechanics, which is whether we can derive the fact that large macroscopic bodies obey Newton's laws, while respecting the assertion that their precise microscopic states may be unique, and hence not part of a large ensemble.   A list of open questions is the substance of section VI, and the  conclusions are stated in section VII.

\section{The real ensemble formulation of quantum mechanics}

\subsection{Kinematics and dynamics of individual systems}

The hypotheses we enunciated above become a formulation and interpretation of quantum mechanics, when we give them a precise instantiation.

\begin{itemize}

\item{}{\bf Kinematics: description of individual states} The state of an individual microscopic system, ${\cal S}_I$  consists of the pair of  beables, 
$( a_I, e^{\imath \phi_I})$

\item{}{\bf The ensemble of similarly prepared states.}  This system is one of $N$ similarly constituted systems in the universe, which have been prepared in the same state and are subject to the same external forces
as they evolve.  These are labeled by $I=1,...,N$.  The state of the  ensemble is specified by the collection of pairs, $\{ (a_I (t) , e^{\imath \phi^I (t)} )\}$.  

\item{}{\bf Ensemble state variables.} The individual system evolves partly by a stochastic process.  Because of this, an observer
studying a particular member of the ensemble, cannot predict with certainty which beables she will measure if she makes a measurement at a later time.
She can predict probabilities for different beables to be observed, which are derived from relative frequencies for the states in the ensemble. 
The relative frequencies 
$n_a (t) $  are defined to be  the number of systems in the ensemble which have beable value $a$ at time $t$.  These
are normalized to $\sum_a n_a = N$.   We will also write $a_I$ for the state of the $I$'th system and $n_J=n_{a_J}$ for the number of copies in the ensemble of the beables of system ${\cal S}^J$.

\item{}{\bf Dynamics of individual systems}  There are two modes of evolution of the beables of a system.  

{\bf Stochastic evolution rule. }There is a stochastic evolution by means of which 
 the system ${\cal S}^I$ can copy the beables of the system ${\cal S}^J$.  

The rate by which system $I$ copies the beables of system $J$ is assumed to be of the form
\f
P( I \mbox{copy} J)  = F (n_I, \phi_I, n_J, \phi_J , a_I, a_J )
\label{copy1}
\ff 
When this happens the properties of the system ${\cal S}^I$ inherits the properties of system ${\cal S}^J$
so that
\f
a^I \rightarrow a^J, \ \ \ \ \ e^{\imath \phi_I} \rightarrow e^{\imath \phi_J}
\ff

We note that the rate a system $I$ copies the state of system $J$ is determined entirely by the 
beables of the two systems
\f
P (I \mbox{copy} J) |_{b=a_J ,a=a_I} = F (n_{a_I} , \phi_{a_I}, n_{a_J}, \phi_{a_J} , a, b ) 
= F (n_{a_I} , \phi_{a_I}, n_{a_J}, \phi_{a_J} )_{a b}
\ff
This defines the rates of copying $F (n_{a_I} , \phi_{a_I}, n_{a_J}, \phi_{a_J} )_{a b}$ as functions of the beables. We note that by
definition the components of $F_{ab}$ must be all positive. 

{\bf Continuous evolution rule.}  When this mixing up or copying of the individual states does not happen, the 
phase evolves continuously in a way that depends on the ensemble.  This must have the general form
\f
\dot{\phi}^I= \sum_J  G(n_I, \phi_I, n_J, \phi_J  , a_I, a_J )
\label{phidot1}
\ff

\item{Evolution of the occupation numbers  $n_a$}

We define the occupation numbers, $n_a$, to be the number of members of the ensemble in state $a$.   They evolve as follows
\begin{eqnarray}
\dot{n}_a &=& \sum_{I}\sum_{J\neq I} \delta_{aa_J}(1-\delta_{aa_I})
\left [
P( I \mbox{copy}J ) - P( J \mbox{copy }I )\right ]
\nonumber \\ 
&=& \sum_{I}\sum_{J\neq I} \sum_{b\neq a }\delta_{aa_J} \delta_{b a_I}
\left [
P( I \mbox{copy} J ) - P( J \mbox{copy }I )\right ]
\nonumber \\ 
&=& \sum_{b\neq a } n_b n_a \left [
F_{ab}  - F_{ba}\right ]
\end{eqnarray}

\item{Evolution of the probability densities}

From this we can write down a law for  the evolution of the probability densities, defined by
\f
\rho_a = \frac{n_a}{N}
\ff
These must evolve as\cite{Vink}
\f
\dot{\rho}_a = \sum_{b \neq a}\left (  \rho_b T_{b \rightarrow a} - \rho_a  T_{a \rightarrow b}  \right ) 
\label{rhodotgeneral}
\ff
where $T_{b \rightarrow a}$ are transition rates.

From (\ref{copy1}) above we see that
\f
T_{b \rightarrow a } 
= F (n_{a_I} , \phi_{a_I}, n_{a_J}, \phi_{a_J} )_{a b} n_a
\ff
This is because the probability to copy a beable value $a$ will be proportional to how many members of the ensemble presently have that value.

{\bf Phase alignment.}  There is a specialization of the evolution rules which we will have to make to derive quantum mechanics from this general framework.
This is that 
\f
\phi^I = \phi_{a_I}
\ff
ie the phases are functions of the variables $a_I$.  This will be called phase alignment.  This is a stable condition, because once set as an initial condition it is preserved by
the evolution rule (\ref{phidot1}).  This is because we have then 
\begin{eqnarray}
\dot{\phi}^I&=& \sum_J  G(n_{a_I}, \phi_{a_I}, n_{a_J}, \phi_{a_J}  , a_I, a_J ) 
\nonumber \\
&=& \sum_b n_{b}  G(n_{a_I}, \phi_{a_I}, n_{a_J}, \phi_{a_J}  , a_I, a_J ) 
\nonumber \\
&=&
\sum_b  G^\prime (n_{a_I}, \phi_{a_I}, n_{a_J}, \phi_{a_J}  , a_I, a_J ) 
\label{phidotnew1}
\end{eqnarray}
This implies that 
\f
\dot{\phi}_a = \sum_b  G^\prime (n_{a}, \phi_{a}, n_{b}, \phi_{b} )_{ab}. 
\label{phidotnew2}
\ff
where $G^\prime (n_{a}, \phi_{a}, n_{b}, \phi_{b} )_{ab} = n_{b}  G(n_{a_I}, \phi_{a_I}, n_{a_J}, \phi_{a_J}  , a_I, a_J )$.

In section V we will describe and study a more general evolution law has solutions which achieve phase alignment, but in this and the next section we assume the phases have been aligned initially.

\end{itemize}

\subsection{Restrictions on the evolution rules}

We can introduce some physical considerations which will allow us to restrict the form of $F$ and $G$. 

\subsubsection{Good large $N$ limit}

First, we do not have any evidence the probabilities for quantum states to evolve depend on the size of the ensemble of
similarly prepared systems.  So we require that  $T_{b \rightarrow a} $ and $G^\prime$ depend on ratios $\frac{n_I}{n_J} $.  We can also posit that only relative phases 
are relevant, so that $T_{b \rightarrow a} $ and $G^\prime$ depend on $e^{\imath (\phi_I -\phi_J)}$.  
These together give us
\f
F (n_I, \phi_I, n_J, \phi_J )_{a b} n_a= F^\prime  (\frac{n_I}{ n_J},  e^{\imath (\phi_{a_I} -\phi_{a_J})})_{a b} 
\ff
and similarly for $G^\prime$.  
\f
G^\prime (n_{a}, \phi_{a}, n_{b}, \phi_{b} )_{ab} = G^\prime (\frac{n_{a}}{n_{b}} ,  e^{\imath (\phi_a -\phi_b)} )_{ab} 
\ff

These equations assume all the $n_a >>1$.  There can be additional terms that go away in the limit $n_a >>1$

\subsubsection{Time reversal invariance}

It is easy to see that these forms are constrained by time reversal invariance.

To see the implications of this let us consider an ansatz, which will be sufficient to recover quantum theory.
\f
 F^\prime (\frac{n_a}{ n_b},  e^{\imath (\phi_{a_I} -\phi_{a_J})})_{a b} = \left ( \frac{n_a}{ n_b} \right )^q {\cal R}(e^{\imath (\phi_{a_I} -\phi_{a_J})})_{a b}
 \label{ansatz1}
\ff
\f
G^\prime (\frac{n_{a}}{n_{b}} ,  e^{\imath (\phi_a -\phi_b)} )_{ab} = \left ( \frac{n_a}{ n_b} \right )^r {\cal U }(e^{\imath (\phi_{a} -\phi_{b})})_{a b}
\label{ansatz2}
\ff

Note that ${\cal R}(e^{\imath (\phi_{a_I} -\phi_{a_J})})_{a b}$ must be positive.

We have then, because $\frac{n_a}{ n_b}=\frac{\rho_a}{ \rho_b}$, 
\f
\dot{\rho}_a = \sum_b  \left ( (\frac{\rho_a}{ \rho_b} )^q \rho_b  {\cal R}(e^{\imath (\phi_{a} -\phi_{b})})_{a b} - 
(\frac{\rho_b}{ \rho_a} )^q \rho_a  {\cal R}(e^{\imath (\phi_{b} -\phi_{a})})_{ba} \right ) 
\ff

{\bf Time reversal}  sends $t \rightarrow -t$ but $\rho_a \rightarrow \rho_a$. Let us suppose it 
also send  $\phi_a \rightarrow \tilde{\phi}_a $.   Then we have under time reversal
\f
\dot{\rho}_a \rightarrow - \dot{\rho}_a = \sum_b  \left ( (\frac{\rho_a}{ \rho_b} )^q \rho_b  {\cal R}(e^{\imath (\tilde{\phi}_{a} -\tilde{\phi}_{b})})_{a b} - 
(\frac{\rho_b}{ \rho_a} )^q \rho_a  {\cal R}(e^{\imath (\tilde{\phi}_{b} -\tilde{\phi}_{a})})_{ba} \right ) 
\ff
We have time reversal invariance if this returns the same equation for $\dot{\rho}_a$.  Recalling the positivity of ${\cal R}(e^{\imath (\phi_{a_I} -\phi_{a_J})})_{a b}$
this can only be solved if $q=\frac{1}{2}$ and $\tilde{\phi}_a = -\phi_a$.  We also have to impose
\f
 {\cal R}(e^{\imath (\phi_{a} -\phi_{b})})_{ba} = {\cal R}(e^{\imath (\phi_{a} -\phi_{b})})_{ab}
\ff
We have then 
\f
\dot{\rho}_a = \sum_b \sqrt{\rho_a \rho_b}  \left (  {\cal R}(e^{\imath (\phi_{a} -\phi_{b})})_{a b} - 
  {\cal R}(e^{\imath (\phi_{b} -\phi_{a})})_{ba} \right ) 
\ff
Insisting on time reversal invariance of $\dot{\phi}$ in (\ref{phidotnew2}) then   implies that
\f
{\cal U} ( z )_{ab} = {\cal U}(\bar{z} )_{ab}.
\label{tr-U}
\ff

However the power $r$ is not fixed by time reversal invariance.

\section{Recovery of the Schroedinger equation}

Let us summarize where we are as a result of our ansatz's plus the imposition of a good large $N$ limit and time reversal invariance.  We have two evolution equations

\begin{eqnarray}
\dot{\rho}_a &=& \sum_{b \neq a} \left ( \rho_a F_{ab}-\rho_b F_{ba}
\right )
\nonumber \\
&=&\sum_b \sqrt{\rho_a \rho_b}  \left (  {\cal R}(e^{\imath (\phi_{a} -\phi_{b})})_{a b} - 
  {\cal R}(e^{\imath (\phi_{b} -\phi_{a})})_{ba} \right ) 
\label{eom1}
\end{eqnarray}
\f
\dot{\phi}_a = \omega_a +  \sum_{b \neq a}  \left ( \frac{n_a}{ n_b} \right )^r {\cal U }(e^{\imath (\phi_{a} -\phi_{b})})_{a b}
\ff
where $\omega_a = {\cal U}_{aa}$ and ${\cal R}_{ab}$ and ${\cal U}_{ab}$ satisfy the properties above.  

We can now expand ${\cal R}_{ab}$ and ${\cal U}_{ab}$ in Fourier series.  
\f
{\cal R}(e^{\imath (\phi_{a} -\phi_{b})})_{ab} = \sum_{n=a}^\infty  R_{ab}^n  \sin^+ ( n (\phi_a -\phi_b ) + \delta_{ab}^n )
\label{simple1}
\ff
\f
{\cal U}(e^{\imath (\phi_{a} -\phi_{b})})_{ab} = \sum_{n=a}^\infty R_{ab}^{\prime n}  \cos ( n (\phi_a -\phi_b ) + \delta_{ab}^{\prime n} )
\label{simple2}
\ff
 To preserve the positivity of $F_{ab}$ and hence ${\cal R}_{ab}$,  we have
\f
 \sin^+ ( \theta)= \left \{ 
\begin{array}{cc}
   \sin ( \theta) &  \mbox{when that is positive}    \\
0   & \mbox{otherwise}      
\end{array}
\right. 
\ff

It is remarkable that just the first term with the further simplifications $R_{ab}^1= R_{ab}^{\prime 1}$ and $\delta_{ab}^{\prime 1}=\delta_{ab}^{1}$
suffices to reproduce quantum mechanics. 
\f
{\cal R}(e^{\imath (\phi_{a} -\phi_{b})})_{ab} =  R_{ab}  \sin^+ ( \phi_a -\phi_b + \delta_{ab})
\label{simple1}
\ff
\f
{\cal U}(e^{\imath (\phi_{a} -\phi_{b})})_{ab} =  R_{ab} \cos ( \phi_a -\phi_b + \delta_{ab})
\label{simple2}
\ff
where, $R_{ab}=R_{ba}$ are positive constants, $\delta_{ab}$ are constant phases which are odd under time reverse and,

This gives us evolution rules
\f
\dot{\rho} =  \sum_{b \neq a} \sqrt{\rho_a\rho_b}  R_{ab}  \sin ( \phi_a -\phi_b + \delta_{ab})
\label{feom1}
\ff
\f
\dot{\phi}_a = \omega_a +  \sum_{b \neq a}  \left ( \frac{n_a}{ n_b} \right )^rR_{ab} \cos ( \phi_a -\phi_b + \delta_{ab})
\label{feom2}
\ff
It is easy to check that with the choice of $r=-\frac{1}{2}$ this reproduces Schroedinger quantum mechanics.  To see this we 
write the general quantum state.  
\f
|\Psi > = 
\left (
\begin{array}{c}
    \sqrt{\rho}_1 e^{-\imath S_1/\hbar}    \\
     \sqrt{\rho}_2 e^{-\imath S_2/\hbar}   \\
     ... \\
     \sqrt{\rho}_M e^{-\imath S_M/\hbar}
\end{array}
\right )
\ff
which clearly is a property of the ensemble and not of an individual physical system.  Here we have defined 
\f
S_a = \hbar \phi_a
\ff

Equations (\ref{feom1}) and (\ref{feom2}) and hence the  evolution rules we posited are then equivalent to evolution via the Schroedinger equation,
\f
\imath \hbar \frac{d\Psi }{dt}= \hat{H} \Psi
\ff
driven by the hermition Hamiltonian
\f
\hat{H}= 
\left(
\begin{array}{ccc}
{\cal E}_1  &    \Delta_{12}  & ... \\
   {\Delta}_{12}^*    &     {\cal E}_2 & ... \\
   ... & ... & ... 
\end{array}
\right)
\label{ham2}
\ff
here we have set
\f
\Delta_{ab}= R_{ab} e^{\imath \delta_{ab}} \hbar
\ff

\subsection{Final form of the evolution rules}

The final form of our evolution rules is
\begin{eqnarray}
{\cal P}(I \mbox{copy} J) & =&   \frac{1}{\sqrt{n_I n_J}}  R_{a_I a_J}  \sin^+ ( \phi_I -\phi_J + \delta_{a_Ia_J })  + {\cal E}_{IJ}
\label{finalform1} \\
\dot{\phi}_I & =& \Omega_I = \omega_{a_I} +  \sum_{J \neq I}  
\frac{1}{\sqrt{n_I n_J}} R_{a_I a_J } \cos ( \phi_I -\phi_J +\delta_{ab}) + {\cal D}_{IJ}
\label{finalform2}
\end{eqnarray}
It must be emphasized that we have derived a correspondence to quantum mechanics only with the proviso that 
$n_a >>1$ and $n_I >> 1$.  When these are not satisfied other terms could come into the evolution rules.  
I have added terms $ {\cal E}_{IJ}$ and ${\cal D}_{IJ}$ to indicate these.

\section{A possible approach to phase alignment}

The elimination of $S(x,t)$ as a function of beable variables, and hence as an ontological entity in its own right, rests on the postulation of a dynamics which achieves phase alignment.  This means that the phases, $\phi_I$  originally assigned independently to each member of the ensemble, become aligned so they depend only on the value of the beable, ie
\f
\phi_I \rightarrow \phi_{a_I}.
\ff  
As we have shown, phase alignment is a fixed point of the dynamics we have postulated in  (\ref{finalform1},\ref{finalform2}).  But is it an attractor?  My investigations of this question have so far been inconclusive.  But this is not the only option.  It may be that the evolution described in (\ref{finalform2}) is an approximation to another dynamical law which achieves phase alignment.  We now describe a possible model for such dynamics.  We shall see that it is easy to show that this model has solutions which achieve phase alignment, but there remains an open question as to the stability of these solutions.  

Consider the following dynamical system, put in Hamiltonian first order form for simplicity.
\f
S= \int dt  \sum_I \left [ \pi^I  ( \dot{\phi}_I -\Omega_I (\phi, n) ) - \frac{1}{2} ( \pi^I)^2  
- \frac{f^2}{2} \sum_{J \in a_I} \sin^2 (\phi_I - \phi_J   )
\right  ]
\label{model1}
\ff
where the model depends on a new parameter, the frequency $f$,  $\Omega_I$ is defined by eq (\ref{finalform2}), and the notation $J \in a_I$ means the subsystem $J$ shares the beable value with $I$.  

We find the momenta are given by 
\f
\pi^I = \dot{\phi}_I -\Omega_I (\phi, n) 
\label{pidef}
\ff
which satisfy the Poisson brackets
\f
\{ \phi_I , \pi^J \} = \delta_I^J
\ff
with the Hamiltonian
\f
H=   \sum_I \left [   \frac{1}{2} ( \pi^I)^2  + \pi^I \Omega_I (\phi, n)
+  \frac{f^2}{2} \sum_{J \in a_I} \sin^2 (\phi_I - \phi_J   )
\right ] 
\ff
The Hamilton equations of motion follow from the Poisson brackets and include (\ref{pidef}) and 
\f
\dot{\pi}^I = -f^2  \sin (\phi_I - \phi_J   ) \cos (\phi_I - \phi_J   ) - \sum_K \pi^K \frac{\partial  \Omega_K (\phi, n)}{\partial  \phi_I}
\label{pidot1}
\ff
Let us take $f$ very large compared to $\omega$ and the components of $R_{ab}$ and consider this evolution in the approximation where the second term can be neglected.
Then we can approximate (\ref{pidot1}) for small phase differences as 
\f
\dot{\pi}^I = -f^2 n_I (\phi_I - \bar{\phi}_{a_I}) + ...
\label{pidot2}
\ff
where 
\f
\bar{\phi}_{a_I} = \frac{1}{n_I} \sum_{J \in a_I} \phi_J
\ff
is the average value of the phases in the subensemble that shares the value of the beable with system $I$.  The Hamiltonian in this approximation is 
\f
H=   \sum_I \left [   \frac{1}{2} ( \pi^I)^2  
+  \frac{f^2}{2} n_I  (\phi_I - \bar{\phi}_{a_I})^2 
\right ] 
\label{ham2}
\ff
In this approximation $\phi_I$ is driven to the minimum of the potential where 
\f
\phi_I = \bar{\phi}_{a_I}
\label{align}
\ff
so the phases align to their average values for each value of the beable.  Once there
we have from (\ref{pidot1}) the full equations of motion.  
\f
\dot{\pi}^I =  - \sum_K \pi^K \frac{\partial  \Omega_K (\phi, n)}{\partial  \phi_I}
\label{pidot3}
\ff
A solution to this is 
\f
\pi_I = \dot{\pi}_I =0   
\label{pisolution}
\ff
This implies 
\f
\dot{\phi}_I = \Omega_I (\phi, n) 
\label{Omega}
\ff
which recovers (\ref{finalform2}), and hence the Schroedinger equation is satisfied.  Hence our model has a degenerate set of zero energy solutions which achieves both phase 
alignment (\ref{align}) and the Schroedinger dynamics.  What we are not, however, able to show is that the solutions (\ref{align},\ref{pisolution},\ref{Omega}) are stable.

We can get a bit more insight by solving the action (\ref{model1}) for $\pi_I$ and writing it in terms of complex variables $z_I=e^{\imath \phi_I}$ which 
satisfy $z_I^* z_I =1$
\begin{eqnarray}
S &=& \int dt  \sum_I \left [ \frac{1}{2} ( \dot{\phi}_I -\Omega_I (\phi, n) )^2 
- \frac{f^2}{2} \sum_{J \in a_I} \sin^2 (\phi_I - \phi_J   )
\right  ]
\nonumber \\
&=& \int dt  \sum_I \left [\frac{1}{2}  ( \dot{z}_I^*  + \imath \Omega_I (z, n) z_I^*  ) ( \dot{z}_I -\imath \Omega_I (z, n) z_I  ) 
- \frac{f^2}{2} \sum_{J \in a_I} \sin^2 (\phi_I - \phi_J   )
\right ] 
\label{model2}
\end{eqnarray}

This shows that the Wallstrom objection\cite{Wallstrom} is not relevent here, because the theory depends on the  phase $z_I=e^{\imath \phi_I}$ rather
than on $\phi_I$ directly\footnote{Thanks to Antony Valentini for suggesting this was the case.}.

Finally, we can note that when  phase alignment is satisfied, the whole system becomes a lagrangian system, with an action principle given by
\f
S= \int dt \sum_a \left ( \rho_a (\dot{\phi}_a -\omega_a ) - \sum_{b \neq a} \sqrt{\rho_a \rho_b} R_{ab} \cos (\phi_a -\phi_b +\delta_{ab} )
\right )
\label{quantumaction}
\ff
This suggests that the $\rho_a$ and $\phi_a$ are conjugate quantities in the phase of the more general theory in which phase alignment is satisfied.

%We also see that the $\Omega_I$ can be interpreted as a one dimensional connection.  The dynamics described above is actually invariant undera time dependent gauge transformation
%\f
%z_I \rightarrow z_I^\prime = z_I e^{\imath \lambda (t)},  \ \ \ \ \ \  \Omega_I \rightarrow \Omega_I^\prime =  \Omega_I - \dot{\lambda}
%\ff
%Given the form of $\Omega_I$ defined by (\ref{finalform2}) this is satisfied if the frequencies transform as
%\f
%\omega_a \rightarrow \omega_a^\prime =  \omega_a - \dot{\lambda}
%\ff
%Hence, only frequency differences, such as $\omega_a -\omega_b$ are physically meaningful.  

\section{The classical limit}  

Once the  conditions are met which are required to derive quantum mechanics, one can continue from there and consider the effect of taking $\hbar \rightarrow 0$.  This should allow us to recover classical mechanics as a limit of quantum mechanics, in the usual way.  But notice that the same conditions we require to get quantum mechanics, which are  large numbers of copies and large occupation numbers, are needed to recover classical mechanics through this route.   This raises the question of whether the theory described here can account for the fact that large macroscopic bodies obey classical dynamics,  when we assert that they do not obey quantum mechanics.  Can we still derive the classical dynamics of large bodies, while still respecting the distinction that the exact quantum states of macroscopic bodies will often be unique?  The following argument shows that it can.  

To show this we can  start from the action principle (\ref{quantumaction}).  Let us consider a simple model of the translational degrees of freedom of the atoms in a body in one dimension, given by a one dimensional array of
sites, with periodic boundary conditions, with $a=1,...P$ labeling the sites.  Let us multiply  (\ref{quantumaction}) by $\hbar$ to define an action ${\cal S}$.  We also can define
the energy $E_a = \hbar \omega_a$, and  the Hamilton-Jacobi function
$S_a=\hbar \phi_a$.  We want to construct a coarse grained model of a macroscopic body so we choose the transition rates to give nearest neighbor interactions, defined with lattice
spaceing $\bar{a}$, 
\f
R_{ab}=  \frac{\hbar}{2m \bar{a}^2} \left ( \delta_{a b+1}+  \delta_{a b-1}
\right )
\ff
We define the potential energy to be
\f
V(a) = E_a + \frac{\hbar^2}{ma^2}
\ff
The action (\ref{quantumaction}) then becomes
\f
{\cal S} = \int dt \sum_a  \rho_a  \left  (\dot{S} +  \frac{1}{2m} ( \partial_x S )^2- V(a) - V_Q + O(\bar{a})
\right  )
\label{calS}
\ff
where the quantum potential is
\f
V_Q = \frac{\hbar^2}{2m}\frac{\nabla^2 \sqrt{\rho} }{\sqrt{\rho}}
\ff
Neglecting the quantum potential or, equivalently, taking $\hbar \rightarrow 0$, we have the following equations of motion
\begin{eqnarray}
\dot{\rho} = \frac{1}{m}  \partial_x (\rho \partial_x S)
\label{probability}
\\
\dot{S}=  - \frac{1}{2m} ( \partial_x S )^2+  V(a)
\label{HJ}
\end{eqnarray}
We recognize (\ref{probability}) as the conservation of probability, with current velocity $v= \frac{1}{m}\partial_x S$, and (\ref{HJ}) as the Hamilton-Jacobi equation. 
%(CHECK SIGNS)  
Thus, we recover an ensemble
of classical systems obeying the Hamilton-Jacobi equation. 

Note that if classical mechanics is construed to be an approximation to quantum dynamics, and the latter is a probabilistic theory of real ensembles, then so must be the former.  That is why we derive classical mechanics in the form of an ensemble of systems whose probabilities evolve in a way that is driven by the Hamilton-Jacobi equation. 

There appears to be a puzzle here.  It seems that an ensemble is required to derive classical mechanics as an approximation to the copy dynamics proscribed by (\ref{finalform1},\ref{finalform2}).  But we have argued that macroscopic bodies have distinct quantum states.  And yet, the derivation of classical dynamics depends on the beable occupation numbers being large. That is a consequence of the fact that we derived classical mechanics as an approximation to quantum mechanics, and therefor require the same conditions for its validity.   Is there a contradiction here?

The resolution of this apparent puzzle is that we can derive the classical description of motion from a model, which is a coarse grained description of the microscopic beables.  Because beables really exist, there can be an exact or fine grained description in terms of beables that is unique and, at the same time, an equally valid coarse grained beables description in which the beable occupation numbers are large.  We can use the latter approximation to study the coarse grained motion of the atoms in the body.  All we have to do is show that beables representing the coarse grained translational states of individual atoms in a macroscopic body satisfy Newton's laws.  It then follows that the centre of mass does as well.  To accomplish this all we need is that the atoms can be described in terms of beables in such a way that they are in ensembles with large occupation numbers.  To do this, we can employ coarse grained beables, which is the occupation numbers of boxes which are large in units of the atomic spacing. 

%There can be no objection to using both fine grained and coarse grained beables in the description of the same body.  Beables being beables, if it makes sense to have a beable which locates the position of an atom on some scale, it must also makes sense to define a beable that measures how many atoms are in a box of dimensions of a larger scale.  
But if we choose the coarse graining sufficiently coarse so that there are many atoms of the body in each box, we are in the domain of large occupation numbers, just from the atoms contained in that macroscopic body.  We can then use the ensemble which is at hand, which is that consisting of the atoms in the body itself.    That means that the copy dynamics can work within the atoms of the body, when we restrict attention to the beables that represent a coarse grained measure of translational motion.
 To do this we consider the above to be a coarse grained model of an ensemble of atoms making up a body and we take the classical limit for the motion of each atom. 

There may be larger ensembles that our atoms are a part of, but all that is needed for our purposes is that there be at least one.  So long as there is an ensemble in which the occupation numbers are large we will derive quantum mechanics, whether that is a subensemble of a larger ensemble or not. 

While we have to choose $\bar{a}$ so the occupation numbers, $n_a$ are large, the validity of the semiclassical approximation requires also that the wavelengths are long, so we can neglect terms of order $\hbar$, particularly the quantum potential.  Hence, we choose 
 the lattice spacing so that
\f
n_a >>1,  \ \ \ \ \ \ \ \hbar^{-1} \bar{a}  \partial_x S << 1
\ff
In this approximation,  $\rho$ describes the ensemble of particles that make up the body, each of which propagates classically.  Thus, the centre of mass of the body also behaves classically.  So, under this  set of assumptions, we have recovered the fact that the centre of mass of large bodies made of many atoms propagates according to classical dynamics.  

%It might seem curious that we have relied on an approximate ensemble description of the translational degrees of freedom in the atom, while asserting that were we to descibe the precise fine grained quantum many body state, it might very well be unique.  
What we did is completely consistent with the principles this approach to quantum mechanics is based on, both in the use of beables and the insistence that all ensembles we invoke are physically real. 
But there is a a deeper level of explanation missing, which would be something analogous to a renormalization group calculation that connects the two levels of description.  More ambitiously, if we knew more about the fundamental theory, which we assert quantum mechanics is an approximation to for small subsystems of the universe, we might be able to both understand the dynamics of unique systems in the universe and justify the derivation of
the copy rule when applied to coarse grained descriptions of their beables.  What we can say at this stage is that   the use of coarse grained models like this is ubiquitous in condensed matter physics and experience shows such models usually succeed when they capture the coarse grained properties of interest in an experiment.

\section{Issues that require more investigation}

We have seen that the hypotheses introduced here have a simple realization which reproduces quantum mechanics.  Nonetheless, as with any novel idea, there
are issues which will require more thought.

\begin{itemize}

\item{}What exactly defines the ensemble that corresponds to the quantum state?  We need it to correspond to the ensemble of systems made from the same constituents, subject to the same forces, that also share the same pure quantum state.  Is there a precise characterization of these ensembles that does not refer to the concept of quantum state?  Does it suffice to say that these systems have the same constituents, preparation and environment?  

%I  have roughly asserted this thiincludes the same constituents, and environment, these go into the definition of the hamiltonian.  We also have required that members of the ensemble have the same preparation, ie same initial quantum state.

Such a characterization of the ensembles only makes sense in a context in which quantum mechanics is asserted to be an approximation to a different cosmological theory. The use of macrosystems to initialize and define preparations of microsystems as a primitive notion has in common with BohrÕs viewpoint that quantum physics requires a distinction between micro and macro systems.  This demands that there be some more fundamental theory that quantum theory approximates for small subsystems of a universe.

\item{}A related issue concerns the relationship between different coarse grainings of the beables used to provide the ensembles from which a quantum theory may be derived.  As we have seen in the discussion of the classical limit, a system such as a macroscopic body whose fine grained description is unique may be coarse grained to yield an ensemble. The copy rule can be applied to different coarse grained models of the same system yielding different quantum mechanical models.  This need not be a conceptual problem, so long as we take the view that quantum mechanics is always an approximation to a deeper theory.  This accords with much of the practice of quantum field theory and statistical physics, which is to regard all the theories in common use as effective theories which are based on some degree of coarse graining of the degrees of freedom.  Because nothing on the derivation of the Schrodinger equation depends on the size of the ensemble, apart from the requirement that all the beable occupation numbers are large, different models, based on different coarse grainings, will lead to different quantum mechanical descriptions, which are presumably related themselves by coarse graining. But there are two very interesting questions for further investigation here.  First, can we work out the precise relationship between coarse graining the dynamics described here and coarse graining the quantum dynamics?  Second, could there be real observable effects coming from corrections to quantum physics that will depend on the size of the ensemble?

\item{} What about composite systems?  Equally important, how are hierarchies of composite systems to be treated?  A quark is part of a quantum system which is a proton, it is also part of a nucleus, an atom, a molecule, a quantum gate.  There are ensembles connected with each of these.  Are the beables associated only with the highest level of the hierarchy that is still quantum mechanical, or can a single beable evolve with respect to several systems it is a part of?

This issue is also crucial for understanding if this proposal can resolve classic issues in quantum theory such as entanglement and Wigner's friend.

\item{} Does the ensemble require a preferred simultaneity to define it?  We could embrace this, in common also with deBroglie-Bohm and assert that the world of beables is one with a preferred notion of simultaneity.  Or we could explore the possibility that the ensemble is defined relativistially, for example to refer to all identically constituted and prepared systems in the whole spacetime.  Recent research in general relativity has revealed that there is a preferred notion of simultaneity that may play a key role in simplifying the dynamics of the theory\cite{shape}.

\item{} What picks the beables?
Do the beables change when the system is put through a different filter?
Or is there a single preferred basis, ie momentum space?

\item{} How is the connection between linear operators and observables non-diagonal in the beables established? Presumably as in dBB probabilities computed in a single basis suffice but it would be good to clarify this.

\item{} The mechanism of phase alignment just discussed is ad hoc and can probably be improved on.  In particular, the question of the stability of the solution that leads to phase alignment and Schroedinger dynamics must be investigated.

\item{}The nodes issue.  This is the most serious problem of this list.  Recall that we required for the recovery of quantum mechanics that all
$n_a >> 0$.  This fails at nodes of wavefunctions, which is for $a$ such that $\rho (a)= n_a =0$. It is easy to see that the correspondence between
the rules we so far posited and quantum mechanics also breaks down when there are such beables.  

For suppose in the initial state defined by the preparation $n_a (t=0) =0$ for some $a=a_0$. Then it follows that $n_a (t) =0$ for all time, for there is nothing to copy.  
Indeed: 
\f
\dot{n}_a = \sum_{b \neq a} \left ( n_b n_a F_{ab}-n_a n_b F_{ba}
\right )
= \sum_b \sqrt{n_a n_b} 
 R_{ab}  \sin ( \phi_a -\phi_b + \delta_{ab }) =0
\ff

To get more insight into this situation, we should look at the second time derivative
\begin{eqnarray}
\ddot{n}_a
&=&\sum_b \sqrt{\frac{n_b}{n_a}}\dot{n}_a R_{ab} \sin ( \phi_a -\phi_b + \delta_{ab }) + ...
\nonumber \\
&=&
\sum_{b\neq a} \sum_{c\neq a}  \sqrt{\frac{n_b}{n_a}}\sqrt{n_a n_c}  R_{ab} R_{ac} \sin ( \phi_a -\phi_b + \delta_{ab })
 \sin ( \phi_a -\phi_c + \delta_{ab })
+ ...
\nonumber \\
&=&
\sum_{b\neq a} \sum_{c\neq a}  \sqrt{n_b n_c} R_{ab} R_{ac} \sin ( \phi_a -\phi_b + \delta_{a_Ia_J })
 \sin ( \phi_a -\phi_c + \delta_{ab })
\end{eqnarray}

Notice that in the passage from the second to the third line we multiply and divide by $n_{a_0}=0$.  The conclusion is then incorrect.

There are two kinds of responses we can make to this issue.  

We assumed all $n_I$ and $n_a$ were large in arguing for the form that led to quantum theory.   If this viewpoint is correct that quantum dynamics fails for systems that are in states that are unique in the universe.  As we indicated above, there could be other terms that come in. 

Nonetheless the problem is easy to address also within the current rules.  All that is required is either 
1) require that the basis chosen for the beables is such that no $n_a=0$ or 
2)  add to the universe a small number of spectator states in each possible a so that no $n_a=0$. 
3) Insist on a tiny admixture to every state of a state with all $n_a$ non-vanishing such as the ground state.

\item{}Might deviations from quantum mechanics be observable?  To test this idea we would like to predict phenomena which do not occur in conventional
quantum mechanics.  The nodes issue is a sign that there must be such phenomena.  When a quantum system is large and complex enough that it has accessible states which are likely to have small occupation numbers in the universe, deviations from quantum mechanics can be expected.  We note that it is likely that these violate signal locality, as has been shown to be necessary with a large class of non-local hidden variables theories out of equilibrium\cite{Antony-nonequilibrium}.  It would be interesting to determine if indeed the possibility of faster than light signalling exists in this formulation of quantum mechanics for cases where quantum dynamics breaks down.

\item{}Is the mixing given by the copy rule (\ref{finalform1}) fast enough to account for observations?  Might there be an observable process of relaxation of a single systems outcomes to the ensemble relative frequency and hence to the quantum mechanical probability distributions?  

\item{}What theory is quantum mechanics an approximation to?  The first assumption of this approach is that quantum mechanics is an approximation to a different, cosmological theory, applicable only for small subsystems that come in many copies.  We then can aspire to discover the principles that this novel theory is based on. A first goal will be search for principles which could characterize such a theory that might be testible in experiments where quantum mechanics is expected to fail because the requirement that occupation numbers be large breaks down.  

\end{itemize}

\section{Conclusions}

Here we have proposed a new interpretation of quantum mechanics based on a new concept of the distinction between a microscopic and macroscopic systems. The distinction is that microscopic systems are those that come in vast numbers of copies in the universe, while macroscopic systems are big and complex enough that they are unique.  Only microscopic systems can satisfy the laws of quantum mechanics, because those laws are consequences of the copy dynamics, and these don't act when there are no systems to copy. 

Hence, 
this proposal addresses the question as to why macroscopic systems do not have quantum properties.  It is simply that if a system is sufficiently composite it has so many possible states that it has no copies within the universe.  It is a member of an ensemble of one.  It is simply a fact that there are a vast number of hydrogen atoms in each of the low lying states within the Hubble scale.  But there is only one you, and only one system identical quantum mechanically to your cat Emily.     This implies that quantum mechanics must be an approximation to a cosmological theory which is formulated in different terms.  

As a result of its limited domain of applicability, the proposal we have made here may have striking consequences for experiment, for it proposes a new regime where quantum dynamics should
fail or receive corrections.  Quantum dynamics should fail both for systems that have no copies in the universe and for systems in states that are unique in the universe.  
This leads us to ask whether it is possible to use the technology of quantum computation to produce a device that can be put into unique, coherent quantum states, unlikely to exist anywhere else?

Similarly, we should expect that the dynamics of systems near nodes may be revealing of the underlying dynamics which replaces quantum theory for individual systems. 

More generally, the new distinction we have introduced between microscopic and macroscopic suggests an exploration for a new regime of mesosccopic physics: those systems which are likely to come in a small numbers of copies in the universe.  The study of such systems should reveal evidence for the underlying laws that quantum mechanics approximates.  

This proposal also implicitly addresses speculation by some theoretical cosmologists that the universe comes in an infinite number of copies which contain many exact and inexact  copies of the Earth and each one of us\cite{Brian}.  Within the present proposal, the fact that macroscopic bodies do not appear to satisfy the superposition principle can be taken as evidence that the universe is finite so that we and other macroscopic bodies have no copies.  On the other hand, testing the limits of the applicability of quantum mechanics to mesoscopic systems like quantum circuits may make it possible to do local measurements which could determine whether there are any copies of them in the universe\footnote{I refrain from exploring some obvious science fiction conjectures about this.}. 

A number of queries and issues can be raised concerning this proposal, some of which were discussed above.  These need to be better understood before the proposal made here can be considered to be in final form.

%I close with a parable.  In a park in Toronto there are many of the usual black squirrels, which so far as one can tell are identical.  Then there is occasionally seen a single, white squirrel.  No one has ever seen two white squirrels together so legend has it there is only one.  Tourists and school groups come from far away to try their luck sighting the white squirrel.   If we describe the squirrels as quantum systems, and treat the park as their universe, is there an important difference in how the grey and white squirrels behave, due to only the former being part of a big ensemble?  The idea proposed here is that there is. 

\section*{ACKNOWLEDGEMENTS}

I am  grateful to Julian Barbour,  Saint Clair Cemin, Cohl Furey, Lucien Hardy, Adrian Kent, Jaron Lanier, Michael Neilsen, Carlo Rovelli, Rob Spekkens and Antony Valentini  for comments, suggestions on drafts of this paper, and encouragement.  I would like to thank also Antony Valentini for the chance to present this work at the New Frontiers in Quantum Foundations, CUPI 2011 conference and also thank the participants for valuable comments.  This direction of thought was inspired by work in progress with Roberto Mangabeira Unger on the implications of the hypothesis that there is only one universe which must be explanatorily closed.  Research at
Perimeter Institute for Theoretical Physics is supported in part by
the Government of Canada through NSERC and by the Province of
Ontario through MRI.

\end{document}